# Scaling rules for multiphase flow in vertical Venturis

**Meng-Ke ZHAN[1, 2], Cheng-Gang XIE[2], Jian-Jun SHU[1] (✉)**

*1. School of Mechanical & Aerospace Engineering, Nanyang Technological University, 50 Nanyang Avenue, Singapore 639798;*
*2. Schlumberger Oilfield (Singapore) Pte Ltd, Singapore Well Testing Center, 1 Benoi Crescent, Singapore 629986*

**Abstract:** Employing a suitable scaling rule in gas-liquid flow can produce dynamically comparable results, which helps in the development of flow models applicable to a wide range of flow conditions and reduces the carbon footprint; however, matching all dimensionless numbers in gas-liquid flow is a challenge. This study uses a computational fluid dynamics approach to identify key dimensionless numbers that can produce dimensionally equivalent results under a variety of flow conditions in the vertical Venturi of varied sizes. The performance of the scaling rule is evaluated and validated based on experimental measurements in terms of the phase fraction, the Venturi dimensionless pressure drop, and the two-phase discharge coefficient.

✉ Correspondence should be addressed to Jian-Jun SHU. *E-mail address*: mjjshu@ntu.edu.sg

# Nomenclature

**Acronyms**

| | |
|---|---|
| CFD | computational fluid dynamics |
| GVF | gas volume fraction |
| MPFM | multiphase flowmeter |
| MT | mid-throat |
| RMSE | root-mean-square error |
| VI | Venturi inlet |

**Greek Symbols**

| | | |
|---|---|---|
| $\alpha$ | phase holdup/fraction | |
| $\beta$ | Venturi throat to Venturi inlet diameter ratio | |
| $\mu$ | dynamic viscosity | Pa s |
| $\rho$ | density | kg/m$^3$ |
| $\sigma$ | surface tension | N/m |

**Roman Symbols**

| | | |
|---|---|---|
| $A$ | area | m$^2$ |
| $C$ | force coefficient | |
| $D$ | Venturi inlet diameter | m |
| $d$ | Venturi throat diameter | m |
| $\mathrm{E_o}$ | Eötvös number | |
| $\mathrm{\tilde{E}_o}$ | modified Eötvös number | |
| $\vec{F}$ | force | N |
| Fr | Froude number | |
| $g$ | gravitational constant | m/s$^2$ |
| $H$ | height | m |
| $KE$ | kinetic energy | J |
| $\vec{k}$ | direction of gravity | |
| $l$ | length | m |
| $n$ | number of phases or number of cells | |
| $P$ | pressure | Pa |
| Re | Reynolds number | |
| $r$ | ratio | |
| SL | slippage number | |
| $UE$ | gravitational potential energy | J |
| $\vec{V}$ | flow velocity | m/s |
| $W$ | work done | J |

**Subscripts**

| | |
|---|---|
| $D$ | drag (force/coefficient) |
| $g$ | gas phase |
| $gamma$ | (evaluated/measured) over gamma ray |
| $HU$ | (evaluated) using phase holdup/fraction |
| $in$ | horizontal inlet |
| $L$ | lift (force/coefficient) |
| $l$ | liquid phase |
| $m$ | mixture |
| $operating$ | operating condition |
| $s$ | superficial velocity |
| $VF$ | (evaluated) using volume fraction |
| $WL$ | wall lubrication (force) |

# 1 Introduction

The difference in the scale of fluid properties and flow conditions between laboratory and field conditions has been highlighted by many researchers as a major challenge in validating flow models (Al-Sarkh *et al.*, 2016; Farokhpoor *et al.*, 2020). For example, different pipe diameters are used for different flow conditions. This can be seen in the design of a Venturi-based multiphase flowmeter (MPFM), where a selection of the different Venturi sizes is often available to suit the measurement needs of oil and gas wells with a wide range of production (Schlumberger, 2017). In addition, laboratory experiments are typically performed at close to atmospheric pressure with low gas densities compared to field condition, whereas gas densities can be at least an order of magnitude higher under higher pressure. Because the Venturi-based

MPFM uses the Venturi differential pressure and phase fraction (mixture density) measurements to calculate flow, a flow model based solely on laboratory conditions may lead to erroneous flow predictions. Hence, there is a need for flow models to be validated against field condition. Although studies (Tayebi *et al*., 2000; Omebere-Iyari *et al*., 2007; Al-Sarkh *et al*., 2016; Farokhpoor *et al*., 2020) have been conducted on gas-liquid flow under flow condition comparable to field condition, the fabrication cost and carbon footprint of extensive experiments under such conditions are high. Therefore, there is a need to develop a method to correlate and predict the flow property of gas-liquid flow at the different scales of flow conditions.

Unlike single-phase flows where scaling studies based on dynamics similitude are common, scaling studies in gas-liquid flows are more challenging. Gas-liquid flow dynamics depends on many flow conditions and fluid properties, making the selection of meaningful dimensionless groups difficult; however, some recent progress has been made in multiphase-scale studies with promising results. Based on the segregation model (Lockhart and Martinelli, 1949), a pressure escalation rule was developed to predict the pressure gradient and liquid holdup at elevated pressure (Al-Sarkh *et al*., 2016). The scaling rules are validated by experiments and simulations of stratified and annular flow in horizontal pipes. It is worth noting that the prediction results are more consistent with the simulation than the experiment, which is due to the more accurate matching of mass flow rate at the two pressure levels in the simulation. Scaling rules were also established based on the segregation model of stratified and annular flow in horizontal or near-horizontal pipes (Shu *et al*., 1997; Shu, 2003a; 2003b; Farokhpoor *et al*., 2020). The agreement between the available experimental data (Linga and Østvang, 1985; Linga and Hedne, 1987; Hedne, 1988; 1996) and the experimental measurements of their respective scaled counterparts provides a good validation of the scaling rule; however, discrepancies were observed in flows with higher gas superficial velocities.

It is observed that scaling rules derived from dynamics similitude can deliver promising solutions in stratified and annular gas-liquid flow; however, current scaling rules have some limitations. First, some approaches involve assumptions that do not hold in certain circumstances. For example, scaling rules (Al-Sarkh *et al*., 2016; Farokhpoor *et al*., 2020) are based on segregation models and do not consider the effects of gas-liquid interaction. This may be the reason that scaling rules are not applicable to dispersed (and intermittent) flows, where gas-liquid interaction, such as drag, lift, and turbulent dispersion, can be important. Second, scaling rules are validated in steady-state, developed, horizontal, or near-horizontal pipe flows. The development process is often encountered in engineering application. Depending on the application, the flow can also take on different pipe inclinations/geometries. For example, in a Venturi-based MPFM, a concurrent upward (or in some cases downward) gas-liquid flow goes through a vertical Venturi. In addition, a Venturi-based MPFM rarely has a sufficient length of flow development leading to the position where the measurements are taken due to constraints on piping and flowmeter placement (Rammohan *et al*., 2013). Overall, to make scaling rules applicable to a wider range of applications, the application of scaling rules for gas-liquid flow needs to be extended to different flow conditions such as dispersed and intermittent flows, developing flows, and flows with different pipe geometries. This study investigates the effect of matching different dimensionless numbers, including quantities related to gas-liquid interactions, on the phase fraction, dimensionless Venturi differential pressure, and two-phase discharge coefficient in a vertical Venturi in an MPFM downstream of the horizontal entrance length.

## 2 Methodologies

### 2.1 Scaling principle

There are many dimensionless numbers available associated with gas-liquid flow. To select the relevant ones, we start by writing the governing equations for the gas-liquid flow in dimensionless form. Several assumptions were made about the governing equations, including (i) a steady framework is used instead of a transient one (Zhan *et al*., 2022), and (ii) the gas-liquid flow is considered incompressible if the criterion $\frac{\Delta P}{P_{line}} < 5\%$ is met, where $\Delta P$ and $P_{line}$ are the Venturi differential pressure and the line pressure, respectively. In this study, nitrogen gas is used as the gas phase. According to the Aungier–Redlich–Kwong equation of state (Aungier, 1995), the nitrogen density difference between the flow inlet and the throat is less than $5\%$, relative to the throat density, if the $\frac{\Delta P}{P} < 5\%$ is satisfied at line pressure $P \sim 20\,\mathrm{bar}$ and temperature $T \sim 30\,^0C$. The simplified governing equations are as follows:

$$\sum_{q=1}^{n} \alpha_q = 1, \qquad (1)$$

$$\vec{\nabla} \bullet \left(\alpha_q \vec{V}_q\right) = 0, \qquad (2)$$

$$\rho_q \vec{\nabla} \bullet \left(\alpha_q \vec{V}_q^2\right) = -\alpha_q \vec{\nabla} P + \vec{\nabla} \bullet \left[\alpha_q \left(\mu_q + \mu_t\right)\left(\vec{\nabla} \vec{V}_q + \vec{\nabla} \vec{V}_q^T\right)\right] \\ + \alpha_q \rho_q g \vec{k} + \sum_{p=1}^{n} \vec{F}_{pq}, \qquad (3)$$

where $\alpha_q = \frac{v_{sq,in}}{|\vec{V}_q|}$, $v_{sg,in}$, $\vec{V}_q$, and $\rho_q$ are the fraction (holdup), the superficial velocity at the horizontal inlet, the flow velocity, and the density of the phase $q$, respectively, and $P$, $g$, $\vec{k}$, $\vec{F}_{pq}$, $\mu_q$, and $\mu_t$ are the pressure shared by all phases, the gravitational constant, the direction of gravity, the phase interaction force acting on the phase $q$, the shear dynamic viscosity of the phase $q$, and the turbulent dynamic viscosity, respectively. For a gas-liquid two-phase flow, the number of phases is $n = 2$.

Equation 1 is already in dimensionless form. To obtain the dimensionless forms of Equations 2 and 3, we multiply both sides of Equations 2 and 3 by $\dfrac{D}{\rho_{g,in} v_{sg,in}^2}$, where $D$ is the Venturi inlet diameter (see Figure 1) and $\rho_{g,in}$ is the gas density at the horizontal inlet. The dimensionless form of the mass continuity and momentum equations for gas and liquid can be written as:

$$\nabla \bullet (\alpha_q \vec{v}_q) = 0, \qquad (4)$$

$$r_{q,\rho} \nabla \bullet (\alpha_q \vec{v}_q^2) = -\alpha_q \nabla p + \nabla \bullet \left[ \dfrac{\alpha_q r_{q,\rho} r_{q,sv}}{\mathrm{Re}_t} \left( \nabla \vec{v}_q + \nabla \vec{v}_q^T \right) \right] + \dfrac{\alpha_q r_{q,\rho} r_{q,sv}^2}{Fr_q^2} \vec{k} + \sum_{p=1}^{n} \vec{f}_{pq}, \qquad (5)$$

where $r_{q,\rho} = \dfrac{\rho_q}{\rho_{g,in}}$, $r_{q,sv} = \dfrac{v_{sq,in}}{v_{sg,in}}$, $\mathrm{Re}_t = \dfrac{\rho_{q,in} v_{sq,in} D}{\mu_q + \mu_t}$, and $\mathrm{Fr}_{sq} = \dfrac{v_{sq,in}}{\sqrt{gD}}$ are the horizontal inlet density ratio, the superficial velocity ratio, the Reynolds number based on the superficial velocity of the phase $q$, and the Froude number based on the superficial velocity of the phase $q$, respectively. Note that $\mu_t$ is the turbulent dynamic viscosity, which varies with the different turbulence models selected for prediction because turbulence is related to the superficial Reynolds number of the phase $q$, $\mathrm{Re}_{sq} = \dfrac{\rho_{q,in} v_{sq,in} D}{\mu_q}$, in the scaling study. Table 1 gives the expressions for dimensionless coordinates and variables.

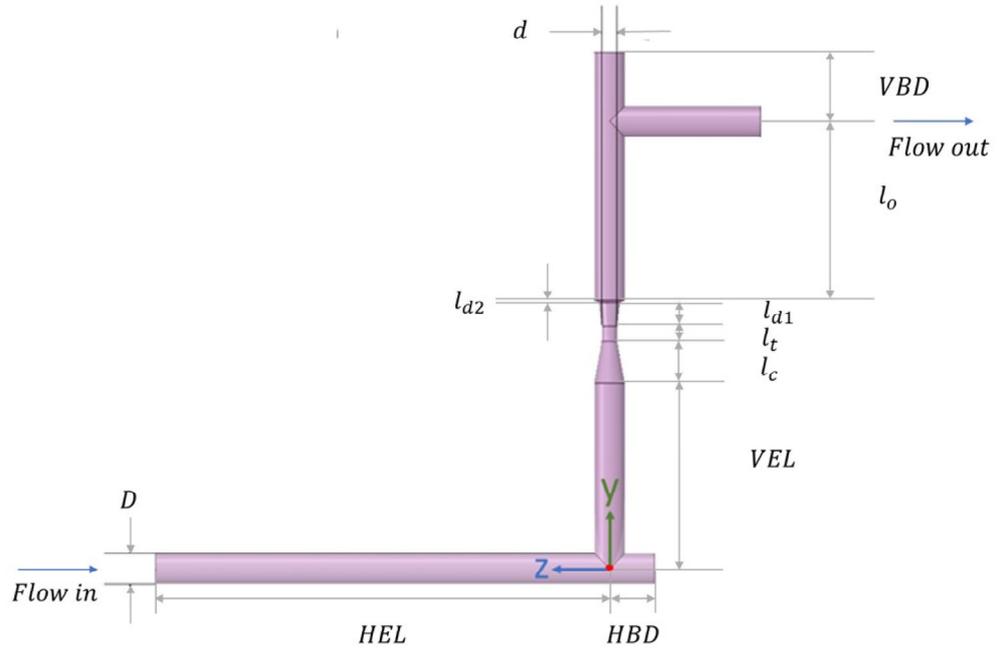

**Figure 1** Geometry of vertical Venturi downstream of horizontal pipe with blind tee

**Table 1** Dimensionless variables in governing equations

| Symbols for dimensionless variables | Dimensional variables |
| --- | --- |
| $\nabla$ | $D\overline{\nabla}$ |
| $\vec{v}_q$ | $\dfrac{1}{v_{sg,in}} \vec{V}_q$ |
| $p$ | $\dfrac{P}{\rho_{g,in} v_{sg,in}^2}$ |
| $\vec{f}$ | $\dfrac{D}{\rho_{g,in} v_{sg,in}^2} \vec{F}$ |

To obtain the same solution to the dimensionless governing Equations 1, 4, and 5 for the different sets of problems, the scaling coordinates, the coefficients of the dimensionless variables, and the corresponding boundary conditions of the dimensionless governing equations need to be held constant. For the scaled coordinates to be the same, the flow domains must be geometrically similar. For a vertical Venturi downstream of a horizontal pipe as shown in Figure 1, this involves maintaining the same geometric similarity of the following length to Venturi inlet diameter ratios: $\beta = \dfrac{d}{D}$, $\dfrac{HEL}{D}$, $\dfrac{HBD}{D}$, $\dfrac{VEL}{D}$, $\dfrac{l_c}{D}$, $\dfrac{l_t}{D}$, $\dfrac{l_{d1}}{D}$, $\dfrac{l_{d2}}{D}$, $\dfrac{l_o}{D}$, and $\dfrac{VBD}{D}$, where $d$, $HEL$, $HBD$, $VEL$, $l_c$, $l_t$, $l_{d1}$, $l_{d2}$, $l_o$, and $VBD$ are the Venturi throat diameter, the horizontal entrance length, the horizontal blind-tee depth, the vertical entrance length, the convergence length, the throat length, the first divergence length, the second divergence length, the outlet pipe length, and the vertical blind-tee depth, respectively.

From the dimensionless governing Equation 5, the coefficients for the dimensionless variables that need to remain unchanged include the horizontal inlet density ratio, $r_{q,\rho}$, the superficial velocity ratio, $r_{q,sv}$, and the Reynolds number, $\text{Re}_{sq}$, based on superficial velocity, and the Froude number, $\text{Fr}_{sq}$, based on superficial velocity. In addition, there are some dimensionless numbers related to the gas-liquid interaction force $\vec{F}_{gl}$. There are many types of gas-liquid interaction forces, such as virtual mass, drag, lift, wall lubrication, and turbulent dispersion forces. Our study identifies drag force $\vec{F}_D$, lift force $\vec{F}_L$, and wall lubrication force $\vec{F}_{WL}$ as the dominant phase interaction forces when simulating gas-liquid flow with negligible compressibility. In order to determine the dimensionless number related to $\vec{F}_{gl}$, after multiplying both sides of $\vec{F}_D$, $\vec{F}_L$, and $\vec{F}_{WL}$ by $\dfrac{D}{\rho_{g,in} v_{sg,in}^2}$, the dimensionless form of the gas-liquid interaction force $\vec{f}_{gl}$, including the dimensionless drag force $\vec{f}_D$, the dimensionless lift force $\vec{f}_L$, and the dimensionless wall lubrication force $\vec{f}_{WL}$, which is given by

$$\vec{f}_D = \frac{3}{4} C_D \frac{D}{d_g} \alpha_g r_{l,\rho} \|\vec{v}_g - \vec{v}_l\| (\vec{v}_g - \vec{v}_l), \quad (6)$$

$$\vec{f}_L = -C_L \alpha_g r_{l,\rho} (\vec{v}_g - \vec{v}_l) \times (\nabla \times \vec{v}_l), \quad (7)$$

$$\vec{f}_{WL} = -C_{WL} D \alpha_g r_{l,\rho} \|\vec{v}_g - \vec{v}_l\|^2 \hat{n}_W, \quad (8)$$

where $C_D$, $C_L$, and $C_{WL}$ are the interphase drag, interphase lift, and wall lubrication force coefficients, respectively. $d_g$ is the Sauter mean diameter (Sauter, 1926) of gas bubble, and $\hat{n}_W$ is the normal component pointing away from the wall. The Tomiyama's drag model (Tomiyama et al., 1998), the Tomiyama's lift model (Tomiyama et al., 2002), and the Frank's wall lubrication model (Frank et al., 2008) are used to compute $C_D$, $C_L$, and $C_{WL}$, respectively. From the models, all three coefficients depend on the Eötvös number, $\text{E}_o = \dfrac{g(\rho_l - \rho_g) d_g^2}{\sigma}$, where $\sigma$ is the surface tension. The Tomiyama's lift model also relies on a modified Eötvös number, $\tilde{\text{E}}_o = \dfrac{g(\rho_l - \rho_g) d_h^2}{\sigma}$, based on the length of the longest axis of the deformable bubble, $d_h$, while the Frank's wall lubrication model relies on the absolute distance from the wall, which makes the wall lubrication force difficult to scale. In general, in order to keep the dimensionless gas-liquid interaction force $\vec{f}_{gl}$ identical, the ratio of the Sauter mean diameter to the Venturi inlet diameter, $r_d = \dfrac{d_g}{D}$, the horizontal inlet liquid-to-gas density ratio, $r_{l,\rho}$, the Eötvös number, $\text{E}_o$, and the modified Eötvös number, $\tilde{\text{E}}_o$, (see Table 2) need to be maintained constant.

Table 2    Gas-liquid dimensionless numbers in the governing equations

| Name of dimensionless number | Symbol | Expression |
| --- | --- | --- |
| Horizontal inlet density ratio | $r_{q,\rho}$ | $\dfrac{\rho_q}{\rho_{g,in}}$ |
| Horizontal inlet superficial velocity ratio | $r_{q,sv}$ | $\dfrac{v_{sq,in}}{v_{sg,in}}$ |
| Reynolds number based on the superficial velocity of the phase $q$ | $\text{Re}_{sq}$ | $\dfrac{\rho_{q,in} v_{sq,in} D}{\mu_q}$ |
| Froude number based on the superficial velocity of the phase $q$ | $\text{Fr}_{sq}$ | $\dfrac{v_{sq,in}}{\sqrt{gD}}$ |
| Eötvös number | $\text{E}_o$ | $\dfrac{g(\rho_l - \rho_g) d_g^2}{\sigma}$ |
| Modified Eötvös number based on the length of the longest axis of the deformable bubble $d_h$ | $\tilde{\text{E}}_o$ | $\dfrac{g(\rho_l - \rho_g) d_h^2}{\sigma}$ |
| Ratio of Sauter mean diameter to Venturi inlet diameter | $r_d$ | $\dfrac{d_g}{D}$ |

| Geometric similarity: length to Venturi inlet diameter ratios | $\beta$ |
|---|---|
| | $\dfrac{d}{D}$ |
| | $\dfrac{HEL}{D}$ |
| | $\dfrac{HBD}{D}$ |
| | $\dfrac{VEL}{D}$ |
| | $\dfrac{l_c}{D}$ |
| | $\dfrac{l_t}{D}$ |
| | $\dfrac{l_{d1}}{D}$ |
| | $\dfrac{l_{d2}}{D}$ |
| | $\dfrac{l_o}{D}$ |
| | $\dfrac{VBD}{D}$ |

For the dimensionless variables to be the same, the dimensionless velocity at the inlet and the pressure at the outlet need to be the same in

$$\left\|\vec{v}_{l,in}\right\| = \frac{r_{l,sv}}{1-\alpha_{g,in}}, \qquad (9)$$

$$\left\|\vec{v}_{g,in}\right\| = \frac{1}{\alpha_{g,in}}, \qquad (10)$$

$$p_{operating} = \frac{P_{operating}}{\rho_{g,in} v_{sg,in}^2}. \qquad (11)$$

In view of the assumption of incompressibility, the effect of pressure changes on the fluid property is negligible. Also, we are interested in the Venturi differential pressure rather than the absolute pressure. Hence, the choice of $P_{operating}$ is arbitrary. Overall, under different flow conditions, a constant dimensionless number is required to obtain the same dynamic solutions as governing Equations 1, 4, and 5 under different flow conditions, are summarized in Table 2.

## 2.2 Computational fluid dynamics model

In this study, a computational fluid dynamics (CFD) approach is used to investigate gas-liquid scaling rules. Unlike experiments, which are subject to system instabilities and measurement uncertainties, the advantage of CFD simulation is that precise flow condition and fluid property can be specified. In addition, a CFD approach can monitor the scaling performance of developing flows as fluid property changes, which is challenging in experiments as measurements need to be made at multiple positions. Furthermore, a CFD approach allows monitoring of dimensionless numbers associated with phase interaction, which is rarely possible experimentally. The Eulerian–Eulerian model is the most comprehensive and widely implemented by many researchers (Shu and Wilks, 1995; Chahed et al., 2003; Reyes-Gutiérrez et al., 2006; Yamoah et al., 2015; Zhang et al., 2019; Acharya and Casimiro, 2020) for solving the governing Equations 1–3, is used for this study. The mixture shear stress transport $k-\omega$ turbulence model (Menter, 1994) is used in the study. Enhanced wall functions are used for near-wall treatment. CFD simulations are performed in a full three-dimensional computational model (see Figure 2) with a polyhedral mesh.

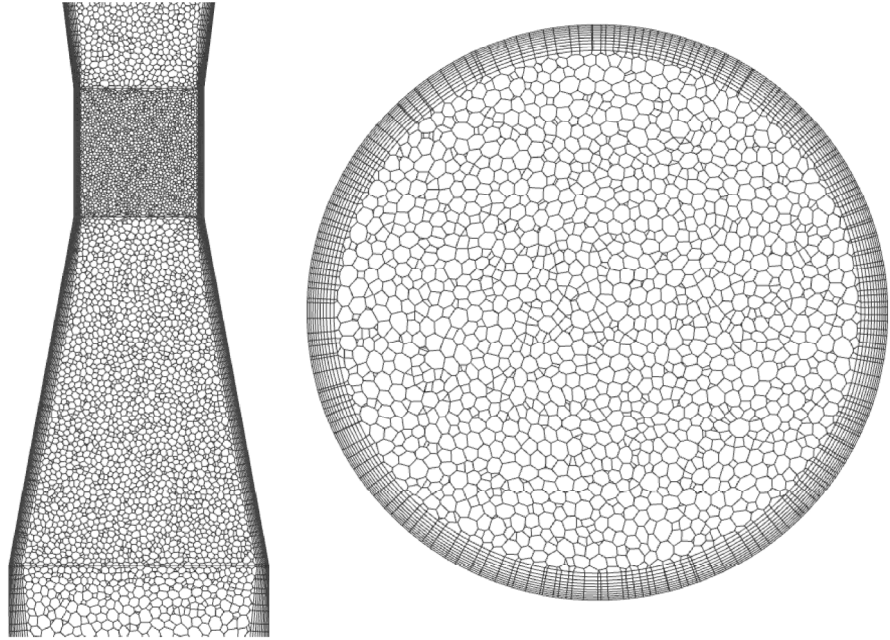

**Figure 2** Front view (left) and cross-sectional view (right) of a polyhedral mesh

This study focuses on the liquid-continuous flow with gas as the dispersed phase. The phase fraction of the dispersed phase is specified at the horizontal inlet, together with the homogeneous inlet velocity. The pressure is specified at the outlet. Sensitivity studies are performed to determine the appropriate size of the Sauter mean diameter, $d_g$, for water-nitrogen and oil-nitrogen flows, respectively. CFD software Ansys Fluent 2021 R1 is used in the study. The phase coupled semi-implicit method for pressure linked equations scheme and the first-order upwind scheme are used for spatial discretization. Interested readers can refer to our earlier work (Zhan *et al*., 2022; 2023) for more details.

CFD simulation results are validated by experimental measurements before being used to study the scaling rules. Mesh-independent results are used in the analysis. Two physical quantities, the Venturi differential pressure, $\Delta P$, and the gamma ray equivalent gas fraction, $\alpha_{g,gamma}$, are used as criteria for validating the simulation results since they are key multiphase measurements for the calculation of multiphase flows. Mesh independence is said to be achieved when $\alpha_{g,gamma}$ changes by less than $0.5\%$ absolute and $\Delta P$ changes by less than $1\%$ relative for an increment of $0.5$ million cells in an observed converging trend. Five test points for water-nitrogen and oil-nitrogen evaluated in a multiphase flow facility are simulated in two Venturi-based vertical MPFMs of sizes S1 and S2 with the Venturi inlet diameters of two inches and three inches, respectively. The line pressure at the test points is about $\sim 20$ bar. The inlet gas volume fraction (GVF) and homogeneous inlet velocity at the horizontal inlet range from $26\%$ to $84\%$, and $1.46$ to $6.29$ m/s. Figures 3(a) and 3(b) show the agreement between the predicted and measured values for the chord-averaged gas fraction, $\alpha_{g,gamma}$, and $\Delta P$, respectively.

3(a)

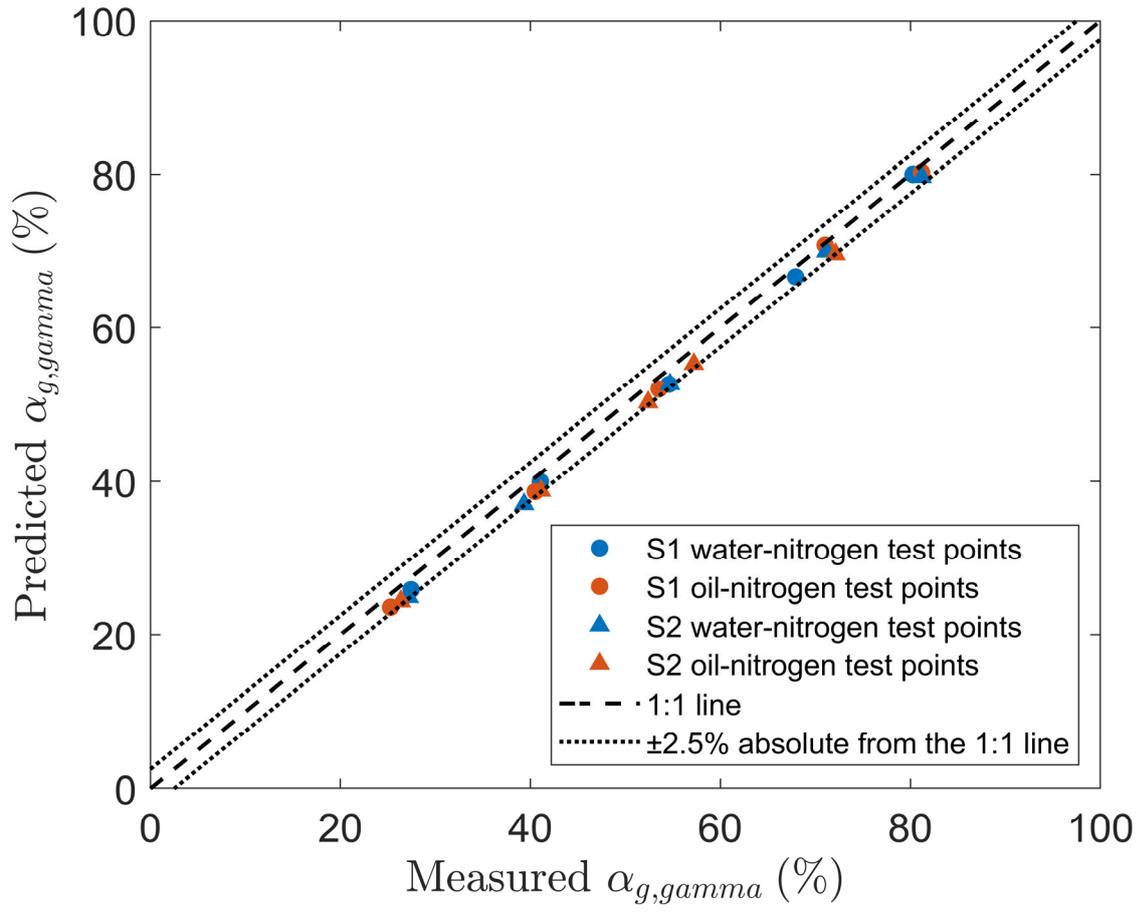

3(b)

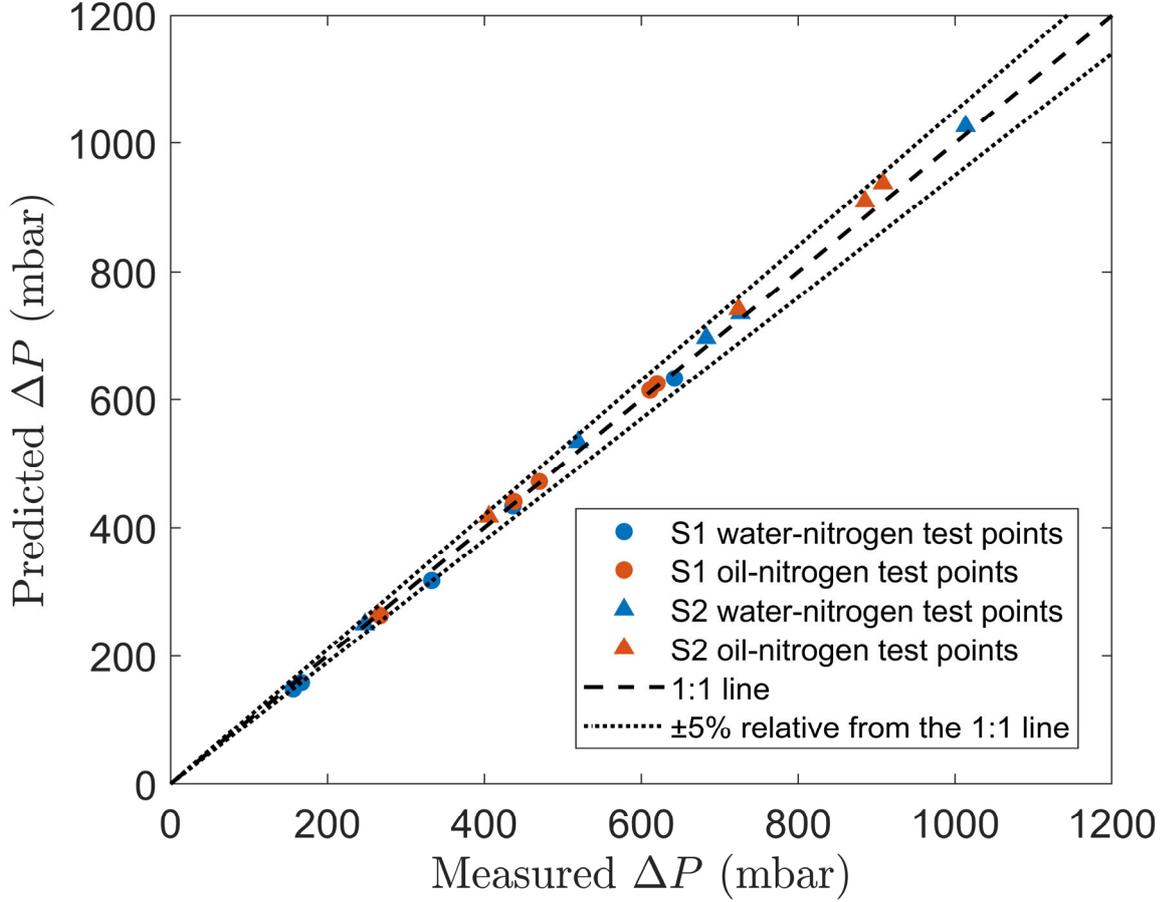

**Figure 3** Computational fluid dynamics-predicted (a) $\alpha_{g,gamma}$ at throat against gamma ray measurement and (b) $\Delta P$ against measurement

For all test points, the absolute difference of $\alpha_{g,gamma}$ and $\Delta P$ obtained from the measurements and CFD simulations are smaller than the 2.5% absolute and 5% relative differences, respectively. This shows that the simulated results agree well with the experimental measurements.

### 2.3 Computational fluid dynamics evaluation matrix

The CFD test matrix with six sets of test points is designed for two purposes: (i) by comparing different sets of test points, the performance of different scaling rules can be evaluated; (ii) the test matrix should cover a range of flow condition and fluid property. It was found that there is an empirical power-law correlation between the slippage number, $SL_p$, and the mixture Froude number, $Fr_m$, as well as the flow patterns such as slender bubbly, slug, slug-annular, churning, wavy-annular, pseudo-slug, and stratified wavy flows clustered along the power-law fitted line (Abdelsalam *et al.*, 2016). The mixture Froude number, $Fr_m$, and the slippage number, $SL_p$, are given by:

$$Fr_m = \frac{v_{sl,in} + v_{sg,in}}{\sqrt{gD}} \sqrt{\frac{\rho_l}{\rho_l - \rho_g}}, \quad (12)$$

$$SL_p = \frac{gD(\rho_{HU} - \rho_{VF})}{\rho_{g,in} v_{sg,in}}, \quad (13)$$

where $\rho_{HU}$ and $\rho_{VF}$ are the mixture density evaluated by phase holdup and (flow rate-based) volume fraction, respectively. In this study, the slippage number $SL_p$ is evaluated using the pipe-averaged phase holdup, and $SL_{gamma}$ is evaluated using the gamma ray equivalent gas fraction.

The flow condition of the test matrix is designed such that the test points cover a range of the Froude number, $Fr_m$, which may correspond to different flow regimes indicted by the correlation. In this study, the range of the mixture Froude number at the throat, $Fr_{m,th}$, is designed to vary from 15 to 25 with a step size of 5 for low- and medium-inlet GVFs 30% and 50% (note that $GVF = \frac{\dot{Q}_g}{\dot{Q}_g + \dot{Q}_l}$, where $\dot{Q}_g$ and $\dot{Q}_l$ are the gas and liquid volumetric flow rates, respectively). For higher inlet GVF 70%, $Fr_{m,th}$ is designed to vary from 30 to 40 at a step of 5. For each test point with a given $Fr_{m,th}$ and inlet GVF, four sets (sets 1 to 4) of gas-liquid (oil-nitrogen or water-nitrogen) flows are simulated using three different gas densities in the computational geometry of the Venturi inlet diameter size S1. Two sets (sets 5 and 6) of water-nitrogen flows with the same liquid property as set 1 are simulated in computational geometry of the Venturi inlet diameter size S2. Whereas sets 1–5 share the same $Fr_{m,th}$, set 6 is obtained by equalizing the $\mathrm{Re}_{l,th}$ of set 6 to that of set 1. This is to study whether the Froude number ($\mathrm{Fr}_m$ or $\mathrm{Fr}_{sq}$) is critical for achieving desirable scaling results. The fluid property and flow condition of the CFD test points are shown in Tables 3 and 4, respectively.

Table 3  Liquid and gas properties for sets 1–6 of test points

| Set | Liquid density ($kg/m^3$) | Gas (nitrogen) density ($kg/m^3$) | Liquid dynamic viscosity ($mPa\,s$) | Gas dynamic viscosity ($mPa\,s$) | Surface tension ($N/m$) |
|---|---|---|---|---|---|
| 1, 5, 6 | 1008.50 (Water) | 24.30 | 1.0 | $1.66\times10^{-2}$ | 0.073 |
| 2 | 799.72 (Oil) | 24.30 | 1.6 | $1.66\times10^{-2}$ | 0.026 |
| 3 | 799.72 (Oil) | 19.27 | 1.6 | $1.66\times10^{-2}$ | 0.026 |
| 4 | 1008.50 (Water) | 30.64 | 1.0 | $1.66\times10^{-2}$ | 0.073 |

Table 4  Flow condition for sets 1–6 of test points

| Group | GVF (%) | $\mathrm{Fr}_{m,th}$ Sets 1–5 | $\mathrm{Fr}_{m,th}$ Set 6 | $v_{m,in}$ (m/s) Set 1 | Set 2 | Set 3 | Set 4 | Set 5 | Set 6 |
|---|---|---|---|---|---|---|---|---|---|
| 1 | 30, 50 | 15 | 9 | 1.98 | 1.98 | 1.98 | 1.98 | 2.32 | 1.45 |
| 2 | 30, 50 | 20 | 12 | 2.65 | 2.64 | 2.65 | 2.64 | 3.09 | 1.93 |
| 3 | 30, 50 | 25 | 16 | 3.31 | 3.30 | 3.31 | 3.30 | 3.87 | 2.42 |
| 4 | 70 | 30 | 19 | 3.97 | 3.96 | 3.97 | 3.96 | 4.64 | 2.90 |
| 5 | 70 | 35 | 22 | 4.63 | 4.62 | 4.63 | 4.62 | 5.41 | 3.39 |
| 6 | 70 | 40 | 25 | 5.29 | 5.27 | 5.29 | 5.27 | 6.19 | 3.87 |

Figure 4 shows the relative magnitudes of the dimensionless numbers of sets 1–6 using normalized dimensionless numbers against set 1. Table 5 shows the range of the dimensionless numbers of test points.

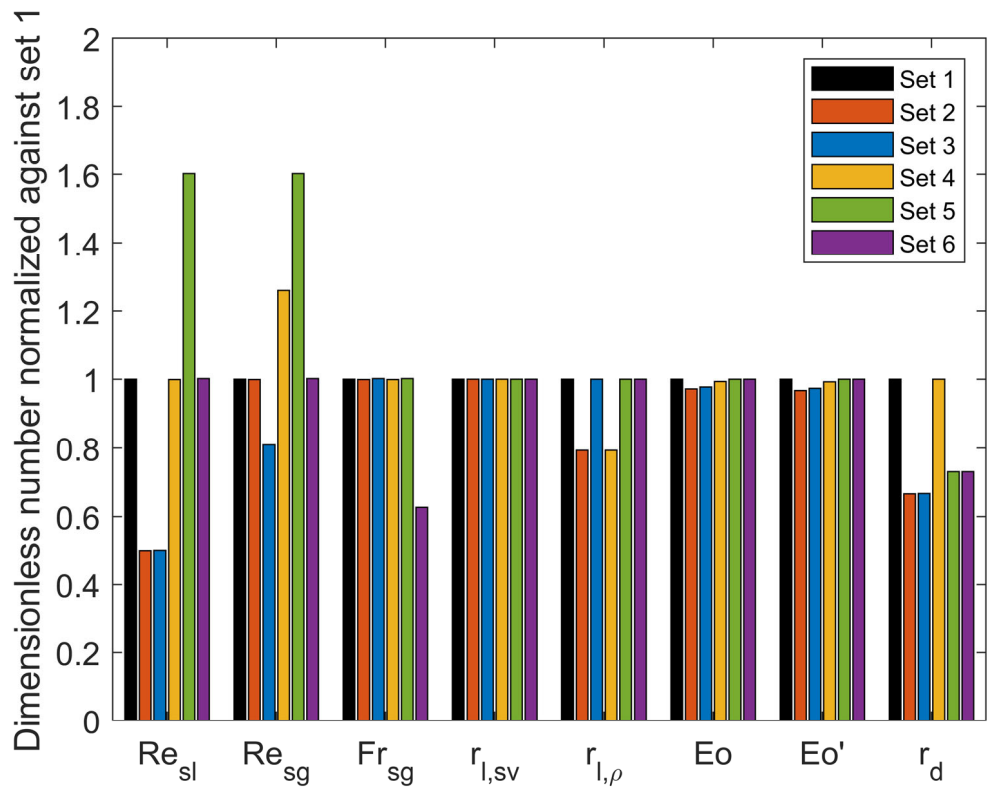

Figure 4  Normalized dimensionless numbers (against set 1) from sets 1–6

Table 5  Range of dimensionless numbers of all test points

| | $\mathrm{Re}_{sl}$ | $\mathrm{Re}_{sg}$ | $\mathrm{Fr}_{sg}$ | $r_{l,sv}$ | $r_{l,\rho}$ | $E_o$ | $\tilde{E}_o$ | $r_d$ |
|---|---|---|---|---|---|---|---|---|
| Minimum | 34485 | 41079 | 0.49 | 0.43 (GVF 70%) | 32.91 | 4.63 | 6.11 | 0.07 |
| Maximum | 215913 | 129469 | 4.89 | 2.33 (GVF 30%) | 41.51 | 4.76 | 6.33 | 0.10 |

## 3 Results and discussion

### 3.1 Phase fraction

#### 3.1.1 Cross-sectional phase fraction

To investigate the effect of scaling rules on the evolution of gas fractions in a vertical Venturi, the simulated cross-sectional gas fractions at the Venturi inlet (VI), $\alpha_{g,\text{VI}}$, and mid-throat (MT), $\alpha_{g,\text{MT}}$, from sets 1–6 at inlet GVFs $30\%$, $50\%$, and $70\%$ are shown in Figure 5.

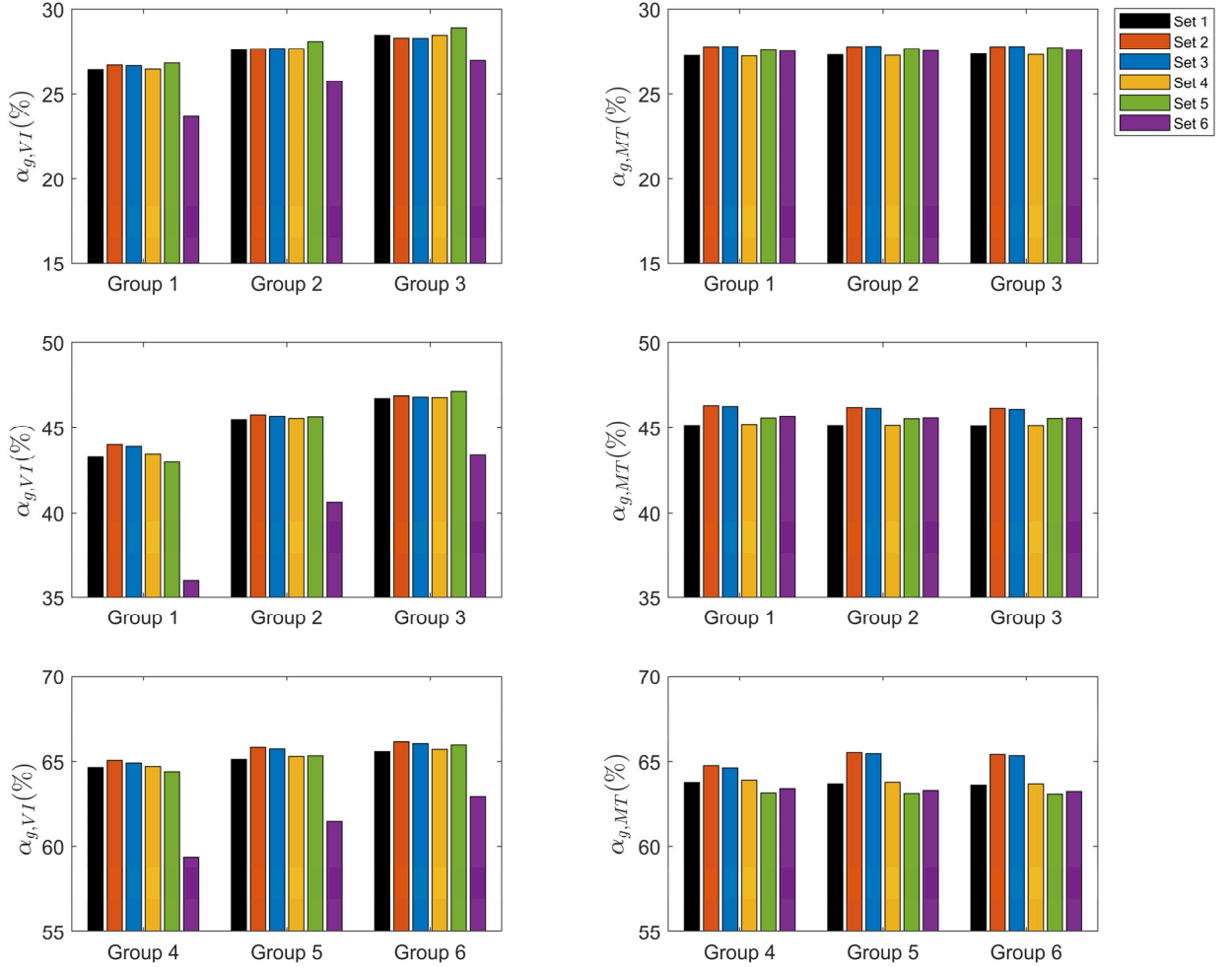

**Figure 5** $\alpha_{g,\text{VI}}$ at the Venturi inlet (left) and $\alpha_{g,\text{MT}}$ at the Venturi mid-throat (right) from sets 1–6 at inlet gas volume fraction $30\%$ (top row), $50\%$ (mid row), and $70\%$ (bottom row) obtained from computational fluid dynamics simulations

From Figure 5, it can be seen that the pair of sets 1 and 4, together with sets 2 and 3, have the best matches, with a difference in $\alpha_{g,\text{VI}}$ of less than $0.17\%$ and a difference in $\alpha_{g,\text{MT}}$ of less than $0.13\%$ for all groups and inlet GVFs. Both pairs share the same $\text{Re}_{sl}$, $\text{Fr}_{sg}$, $r_{l,sv}$, and $r_d$. The $\text{E}_o$ and $\tilde{\text{E}}_o$ within the pair of sets 1 and 4 and the pair of sets 2 and 3 are also relatively close, with a percentage difference of $0.7\%$ (of set 1) and $0.6\%$ (of set 3), respectively, which is less than the percentage difference with the other sets ($\sim 2\%$).

Simulations for sets 5 and 6 are performed in S2. By comparing sets 5 and 6 with the other sets simulated in S1, the meaningful dimensionless numbers for upscaling can be investigated. It can be observed that at VI, $\alpha_{g,\text{VI}}$ from set 6 with a $\text{Fr}_{sg}$ that is lower than the other sets has a significantly lower value with a maximum difference of $7.28\%$ (GVF $50\%$, Group 1) with respect to set 1, which shares the same $\text{Re}_{sl}$, $\text{Re}_{sg}$, $r_{l,sv}$, $r_{l,\rho}$, $\text{E}_o$, and $\tilde{\text{E}}_o$. This may be because set 6 belongs to a different flow regime. The difference in $\alpha_{g,\text{MT}}$ decreases with $\text{Fr}_{sg}$ and increases with inlet GVF. At MT, agreement in $\alpha_{g,\text{MT}}$ between set 6 and the other sets improves from VI, with a maximum difference of $0.52\%$ (GVF $50\%$, Group 1) compared to set 1. At MT, $\text{Fr}_{sg}$ is about $\sim 4$ times that at VI; it is possible that set 6 shares the same flow regime as the other sets at the Venturi throat section.

### 3.1.2 Correlation between slippage number $SL_p$ and mixture Froude number $Fr_m$

To gain a deeper understanding of gas-liquid slippage in a vertical Venturi, Figure 6 shows slippage number $SL_p$ versus the local $Fr_m$ at VI and MT. $SL_p$ is evaluated with density $\rho_{HU}$ calculated from the cross-sectional gas fraction (via Equation 13).

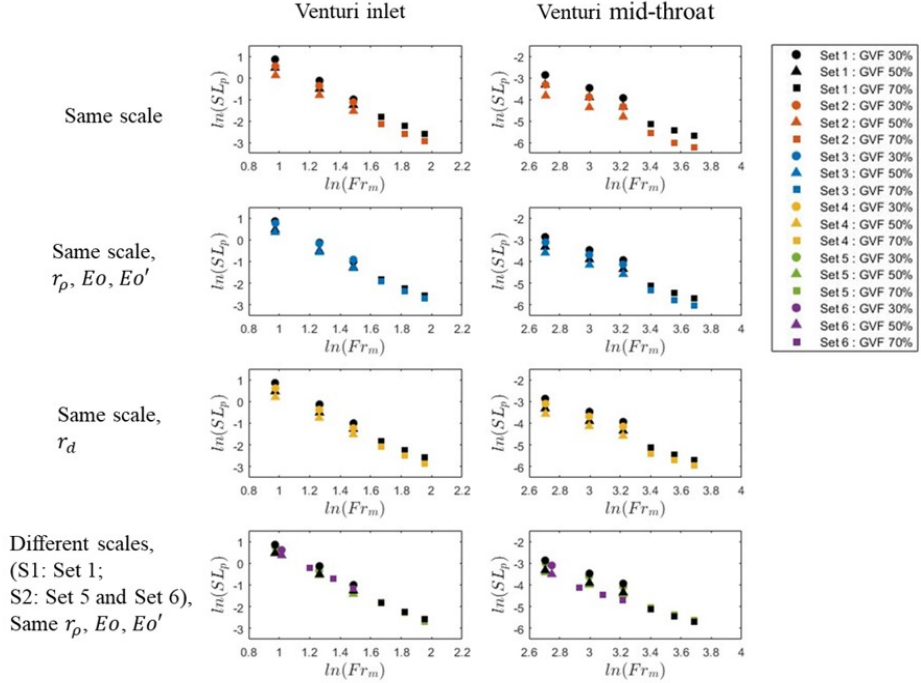

**Figure 6** $\ln(SL_p)$ against $\ln(Fr_m)$ in the Venturi mid-throat at inlet gas volume fraction $30\%$, $50\%$, and $70\%$. Set 1 is used as a reference and compared to set 2 (first row), set 3 (second row), set 4 (third row), set 5, and set 6 (fourth row).

Since each set of test points with different $Fr_{sg}$, $Re_{sl}$, $Re_{sg}$, $r_{l,sv}$ collapses into the same line, the differences in gradient $a$ and offset $b$ are likely due to dimensionless numbers related to fluid property and gas-liquid interaction, such as $r_{l,\rho}$, $E_o$, $\tilde{E}_o$, and $r_d$. It can be seen that at VI, sets 1 and 3 that share the same $r_{l,\rho}$, $E_o$, and $\tilde{E}_o$ within the same scale of the flow domain have the best agreement in $\ln(SL_p)$, with the largest difference of 0.13 ($\frac{SL_{p,set1}}{SL_{p,set3}} = 1.1$). Sets 1 and 2, which do not share any dimensionless numbers related to gas-liquid phase interaction, have the worst agreement, with the largest difference in $\ln(SL_p)$ of 0.39 ($\frac{SL_{p,set1}}{SL_{p,set2}} = 1.5$). In MT, the variance in $\ln(SL_p)$ between different sets increases. Overall, sets 1 and 3 sharing the same $r_{l,\rho}$, $E_o$, and $\tilde{E}_o$, and sets 1 and 4 sharing the same $r_d$ within the same scale of the flow domains have better agreement in $\ln(SL_p)$, with the largest difference in $\ln(SL_p)$ of 0.35 ($\frac{SL_{p,set1}}{SL_{p,set3}} = 1.4$) and 0.25 ($\frac{SL_{p,set1}}{SL_{p,set4}} = 1.3$), respectively. The maximum difference in $\ln(SL_p)$ between sets 1 and 2, which does not share any dimensionless numbers related to gas-liquid phase interaction, and sets 1, 5, and 6, which are different dimensions, are 0.58 ($\frac{SL_{p,set1}}{SL_{p,set5}} = 1.8$) and 0.76 ($\frac{SL_{p,set1}}{SL_{p,set6}} = 2.1$), respectively.

The effect of matching different dimensionless numbers on the coefficients $a$ and $b$ of the best-fit model $\ln(SL_p) = a\ln(Fr_m) + b$ is also investigated. Table 6 shows the coefficients $a$ and $b$ of the best-fit model $\ln(SL_p) = a\ln(Fr_m) + b$ obtained for each set at VI and MT, respectively. The indicators for the goodness of fit including the R-squared value and root-mean-square error (RMSE) of each line-of-best-fit are also shown in Table 6.

**Table 6** For computational fluid dynamics sets 1–6: coefficients $a$ and $b$ of the lines-of-best-fit $\ln(SL_p) = a\ln(Fr_m) + b$ at the Venturi inlet and mid-throat, and the respective fitted R-squared value and root-mean-square error (RMSE)

| Set | Venturi inlet | | | | Venturi mid-throat | | | |
|---|---|---|---|---|---|---|---|---|
| | $a$ | $b$ | R-squared | RMSE | $a$ | $b$ | R-squared | RMSE |
| 1 | −3.38 | 3.93 | 0.98 | 0.17 | −2.81 | 4.62 | 0.94 | 0.24 |
| 2 | −3.43 | 3.71 | 0.98 | 0.21 | −2.83 | 4.26 | 0.92 | 0.30 |
| 3 | −3.42 | 3.91 | 0.98 | 0.20 | −2.81 | 4.42 | 0.93 | 0.30 |
| 4 | −3.40 | 3.70 | 0.98 | 0.18 | −2.77 | 4.29 | 0.93 | 0.29 |
| 5 and 6 | −3.48 | 4.00 | 0.99 | 0.12 | −2.36 | 3.10 | 0.95 | 0.24 |

As can be seen from Table 6, the linear fit between $\ln(\mathrm{SL}_p)$ and $\ln(\mathrm{Fr}_m)$ for all sets at VI and MT is good with the R-squared value $\geq 0.98$ for VI and $\geq 0.92$ for MT. RMSEs for all sets are $\leq 0.21$ for VI and $\leq 0.30$ for MT. At VI, the value of the gradient $a$ of the best-fit at VI is similar for sets 1–4, with the greatest relative difference between sets 1 and 2 of $1.48\%$ (of set 1). For scaling between dimensions S1 and S2, sets 5 and 6 have a $2.96\%$ higher gradient than set 1, which shares the same $r_{l,\rho}$, $\mathrm{E}_o$, and $\tilde{\mathrm{E}}_o$. Within the same dimension S1, pair of sets 1 and 4 and pair of sets 2 and 3 that share the same $r_d$ and similar $\mathrm{E}_o$ and $\tilde{\mathrm{E}}_o$ have a better match in the gradient $a$ between them, compared to the other sets. Within the same dimension S1, the offsets $b$ of the pair of sets 1 and 3 and the pair of sets 2 and 4, which share the same $r_{l,\rho}$, are similar. Note that the average offset $b$ of the sets 1 and 3 is $5.8\%$ higher than that of sets 2 and 4. This may be because that sets 1 and 3 have a $25\%$ higher density ratio $r_{l,\rho}$ than sets 2 and 4, which results in a higher slippage number at any given $\mathrm{Fr}_m$, given the same gradient $a$. For scaling between S1 and S2, sets 5 and 6 have a $1.78\%$ higher offset $b$ than set 1. At MT, it can also be observed that the value of the gradient $a$ of the best-fit for sets 1–4 remains similar (with a maximum $2.17\%$ (of set 4) difference between sets 2 and 4), and the pair with the same $r_{l,\rho}$ has similar offset $b$ are still valid among sets 1–4 test points in S1. For scaling between different dimensions S1 and S2, the best-fit gradient $a$ and offset $b$ for sets 5 and 6 are $16.01\%$ (gradient $a$) and $32.9\%$ (offset $b$) smaller than those for set 1. This indicates that there is a small decrease in slip in S2 than that in S1 for a given increase of $\mathrm{Fr}_m$ at MT. Sets 5 and 6 share the same $r_{l,\rho}$, $\mathrm{E}_o$, and $\tilde{\mathrm{E}}_o$ as set 1. Hence, while up-scaling rule by matching $r_{l,\rho}$, $\mathrm{E}_o$, and $\tilde{\mathrm{E}}_o$ may yield a similar $\ln(\mathrm{SL}_p) - \ln(\mathrm{Fr}_m)$ correlation for straight pipe sections with a certain developing length (*e.g.*, a VI with a $\sim 6D$ vertical upstream developing length), the upscaling rule performs less well for a MT with a short developing length ($0.5D$). Hence, the same pipe sizes must be used to establish the $\ln(\mathrm{SL}_p) - \ln(\mathrm{Fr}_m)$ correlation for slip prediction in MT. With a longer developing length, the greater gas-liquid drag force due to the greater magnitude of the phase velocity difference may reduce the slip (Shu *et al*., 2016; 2017; 2018).

Since the phase fraction is measured by a gamma ray sensor at the throat in the studied Venturi-based MPFM, it is useful to study the slippage number $\mathrm{SL}_{gamma}$ evaluated with $\rho_{HU}$ calculated from the gamma ray equivalent phase fraction from CFD. Figure 7 shows the comparative changes of $\ln(\mathrm{SL}_{gamma})$ against $\ln(\mathrm{Fr}_m)$ for different sets. Table 7 shows the coefficients $a$ and $b$ of the best-fit model $\ln(\mathrm{SL}_{gamma}) = a\ln(\mathrm{Fr}_m) + b$ obtained for each set.

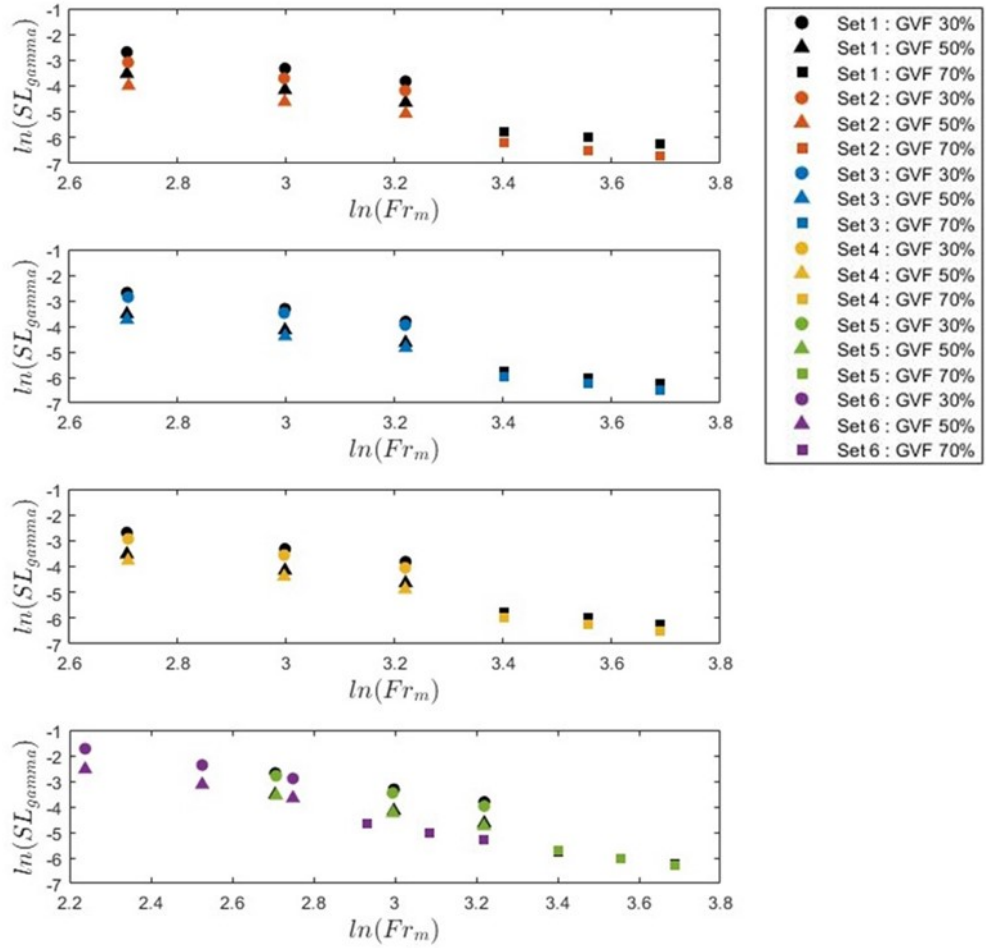

**Figure 7** $\ln(\mathrm{SL}_{gamma})$ against $\ln(\mathrm{Fr}_m)$ in the Venturi mid-throat at inlet gas volume fraction $30\%$, $50\%$, and $70\%$. Set 1 is used as a reference and compared to set 2 (first row), set 3 (second row), set 4 (third row), set 5, and set 6 (fourth row).

**Table 7** For computational fluid dynamics sets 1–6: coefficients $a$ and $b$ of the lines-of-best-fit $\ln(\mathrm{SL}_{gamma}) = a\ln(\mathrm{Fr}_m) + b$, and the respective R-squared and root-mean-square error (RMSE) for each set

| Set | $a$ | $b$ | R-squared | RMSE |
|---|---|---|---|---|
| 1 | −3.42 | 6.38 | 0.86 | 0.84 |
| 2 | −3.48 | 6.14 | 0.85 | 0.54 |
| 3 | −3.48 | 6.34 | 0.85 | 0.54 |
| 4 | −3.44 | 6.17 | 0.86 | 0.51 |
| 5 and 6 | −3.00 | 4.78 | 0.87 | 0.49 |

From Figure 7 and Table 7, it can be seen that the test points in the $\ln(\mathrm{SL}_{gamma})$ versus $\ln(\mathrm{Fr}_m)$ correlation are more dispersed than those in the $\ln(\mathrm{SL}_p)$ versus $\ln(\mathrm{Fr}_m)$ correlation, as evidenced by the lower R-squared and higher RMSE values in the $\ln(\mathrm{SL}_{gamma})$ versus $\ln(\mathrm{Fr}_m)$ correlation for all CFD sets. This may be due to the greater dependence of the directional ($z$-direction) chord-averaged gas fraction on the phase distribution. The flow with identical $\mathrm{SL}_p$ at MT may not have the same $\mathrm{SL}_{gamma}$ if the phase distribution is different; however, the agreement in $\ln(\mathrm{SL}_{gamma})$ between different sets is similar to that in $\ln(\mathrm{SL}_p)$. The pair of sets 1 and 3 that shares the same $r_{l,\rho}$, $\mathrm{E}_o$, and $\tilde{\mathrm{E}}_o$, and the pair of sets 1 and 4 that shares the same $r_d$ within the same scale of the flow domains have better agreement than the pair of sets 1 and 2 that does not share any dimensionless numbers related to gas-liquid phase interaction, and sets 1, 2, and 5 that are of different dimensions. In addition, similar to $\ln(\mathrm{SL}_p)$ versus $\ln(\mathrm{Fr}_m)$ in MT, the set pair with the same $r_d$ and similar $\mathrm{E}_o$ and $\tilde{\mathrm{E}}_o$ has similar gradient $a$ (sets 1 and 4; sets 2 and 3), and the pair with the same $r_{l,\rho}$ has similar offset $b$ (sets 1 and 3; sets 2 and 4) in the dimension S1. For upscaling from S1 to S2, the gradient $a$ and offset $b$ of sets 5 and 6 are $12.28\%$ and $25.08\%$ lower than that in set 1 that shares the same $r_{l,\rho}$, $\mathrm{E}_o$, and $\tilde{\mathrm{E}}_o$. Hence, the coefficient of $\ln(\mathrm{SL}_{gamma})$ versus $\ln(\mathrm{Fr}_m)$ correlation must be identified separately in different dimensions of the flow do-mains.

### 3.2 Dimensionless Venturi differential pressure

In the previous straight pipe scaling study, the pressure drop in the straight pipe was due to friction losses, while the pressure drops across VI and MT was due to the energy conversion of work done by pressure into kinetic energy and gravitational potential energy, as described by the Bernoulli equation below:

$$P_{\text{VI}} + \alpha_{g,\text{VI}}\rho_{g,\text{VI}}gH_{\text{VI}} + \frac{1}{2}\alpha_{g,\text{VI}}\rho_{g,\text{VI}}\left\|\vec{V}_{g,\text{VI}}\right\|^2 + \alpha_{l,\text{VI}}\rho_{l,\text{VI}}gH_{\text{VI}} \\
+ \frac{1}{2}\alpha_{l,\text{VI}}\rho_{l,\text{VI}}\left\|\vec{V}_{l,\text{VI}}\right\|^2 = P_{\text{MT}} + \alpha_{g,\text{MT}}\rho_{g,\text{MT}}gH_{\text{MT}} \\
+ \frac{1}{2}\alpha_{g,\text{MT}}\rho_{g,\text{MT}}\left\|\vec{V}_{g,\text{MT}}\right\|^2 + \alpha_{l,\text{MT}}\rho_{l,\text{MT}}gH_{\text{MT}} \\
+ \frac{1}{2}\alpha_{l,\text{MT}}\rho_{l,\text{MT}}\left\|\vec{V}_{l,\text{MT}}\right\|^2 + \text{Frictional Loss},
\qquad(14)$$

where $H$ is the height. Under the assumption of incompressible flow, no slip, and no friction loss, the dimensionless Venturi differential pressure gradient can be described as

$$\left|\vec{k}\bullet\nabla p\right| = \frac{D(P_{\text{VI}} - P_{\text{MT}})}{\rho_{g,in}v_{sg,in}^2(H_{\text{MT}} - H_{\text{VI}})} \\
= \frac{1 + r_{l,sv}r_{l,\rho}}{\text{Fr}_{sg}^2(1 + r_{l,sv})} + \frac{\left(\dfrac{1}{\beta^4} - 1\right)(1 + r_{l,sv}r_{l,\rho})(1 + r_{l,sv})}{2\left(1 + \dfrac{l_c}{D}\right) + \dfrac{l_t}{D}}, \qquad(15)$$

by understanding the dimensionless numbers, $\text{Fr}_{sg}$, $r_{l,\rho}$, and $r_{l,sv}$, at the horizontal inlet, as well as the dimensionless geometric properties, $\beta$, $\dfrac{l_c}{D}$, and $\dfrac{l_t}{D}$.

### 3.3 Two-phase discharge coefficient

In practice, the Venturi differential pressure is usually larger than that estimated by the Bernoulli equation as additional energy is required to account for the energy loss in the Venturi. Hence, the single-phase discharge coefficient, $C_d$, is introduced to account for the loss. For single-phase flow, $C_d$ increases with $\text{Re}_{sl}$ in a converging trend. To investigate whether the observation hold for two-phase flow, two-phase discharge coefficient, $C_{d,tp}$, is evaluated from CFD simulation results. $C_{d,tp}$ is defined here as

$$C_{d,tp} = \frac{(KE_{\text{MT}} - KE_{\text{VI}}) + (UE_{\text{MT}} - UE_{\text{VI}})}{W_{\text{MT}} - W_{\text{VI}}}, \qquad(16)$$

where $KE$, $UE$, and $W$ are the kinetic energy, the gravitational potential energy, and the work done, respectively. Figure 8 shows the variation of the two-phase discharge coefficient, $C_{d,tp}$, with $\text{Re}_{sl}$.

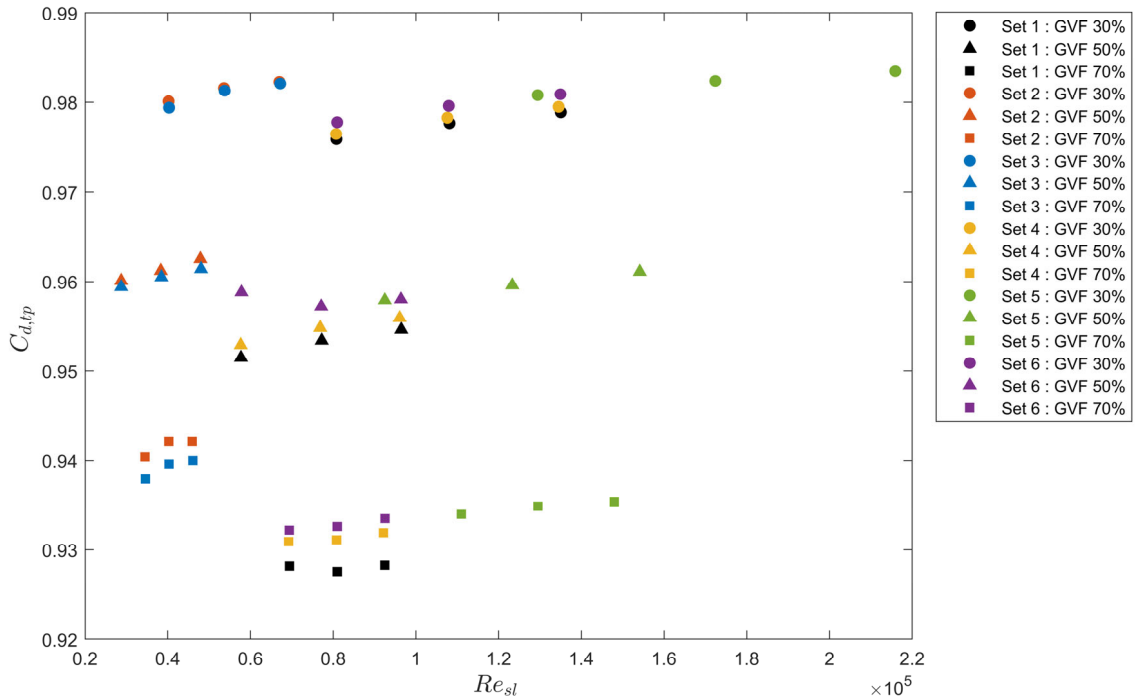

**Figure 8** Variation of $C_{d,tp}$ with $\text{Re}_{sl}$. Circle, triangle, and square represent Groups 1–3 at inlet gas volume fraction (GVF) 30% and 50%, and Groups 4–6 at inlet GVF 70%, respectively.

A few observations can be made from Figure 8. First, there are six clusters of data in the $C_{d,tp}$ versus $\text{Re}_{sl}$ plot. Sets 2 and 3, which share close $\text{E}_o$ and $\tilde{\text{E}}_o$, form three clusters, with the cluster from inlet GVF 30% having the highest $C_{d,tp}$, followed by the clusters with inlet GVF 50% and GVF 70%. Similarly, sets 1, 4–6, which share close $\text{E}_o$ and $\tilde{\text{E}}_o$, form three clusters of inlet GVF 30%, 50%, and 70%, respectively. Within each cluster, $C_{d,tp}$ increases with $\text{Re}_{sl}$ in a converging trend.

Within each cluster, set 2 with lower density contrast $r_{l,\rho}$ has a smaller loss or a higher $C_{d,tp}$ than set 3. The difference increases as inlet GVF increases. Similarly, set 4 with lower $r_{l,\rho}$ also has a smaller loss or a higher $C_{d,tp}$ than set 1. In addition to friction loss between the fluid and the pipe wall, loss may also occur due to gas-liquid interaction. Given dynamically similar flow condition and flow domain S1, lower $r_{l,\rho}$ is likely to result in a more uniform flow, as can be seen from sets 2 and 4 having smaller slip than sets 1 and 3, respectively. Hence, loss due to gas-liquid interaction in sets 2 and 4 is lower than in sets 1 and 3, given the same $\text{Re}_{sl}$.

## 4 Experimental validations

As discussed in Section 2.2, experimental studies of scaling rules often face challenges, including, for example, imperfect matches in flow condition such as inlet GVF and flow velocity. As can be seen from Section 3.1, it is observed that the magnitude of the difference in gas fractions can be as low as a few percent, which is easily within the range of an imperfect match of flow conditions. As a result, it is more meaningful to experimentally verify the correlation (or trend) of the dimensionless number with the gas fraction, the dimensionless Venturi differential pressure, and the two-phase discharge coefficient.

Experiments are performed at SLB (formerly Schlumberger) multiphase flow facility connected to a Venturi-based MPFM and a three-phase separation system. A table referring to the raw experimental data, including line pressure, inlet GVF, velocities, and fluid properties, can be found in Appendix. The phase fraction $\alpha_{g,gamma}$ at MT is measured by a gamma ray sensor, and $\Delta P$ across VI and MT is measured by a Venturi differential pressure sensor installed in the Venturi-based MPFM.

Five sets with six test points each are chosen from two flow loops (sets 1, 2, 4, and 5 in flow loop 1 and set 3 in flow loop 2), such that the fluid properties follow those of sets 1–5, respectively. We do not distinguish between sets 5 and 6 because set 6 with the CFD simulations shares the same fluid property as set 5, but takes into account different $\text{Re}_{sl}$, $\text{Re}_{sg}$, and $\text{Fr}_{sg}$, in the correlation. The minimum and maximum values of the dimensionless numbers for sets 1–5 test points used in the experimental validation normalized against the set 1 maximum are shown in Figure 9. Note that because information on the Sauter mean diameter $d_g$ is not available, $d_g$ used in the CFD simulation is used to compute $\text{E}_o$, $\tilde{\text{E}}_o$, and $r_d$. All dimensionless numbers are evaluated at the horizontal inlet. The range of the dimensionless numbers used in the experimental validation is shown in Table 8.

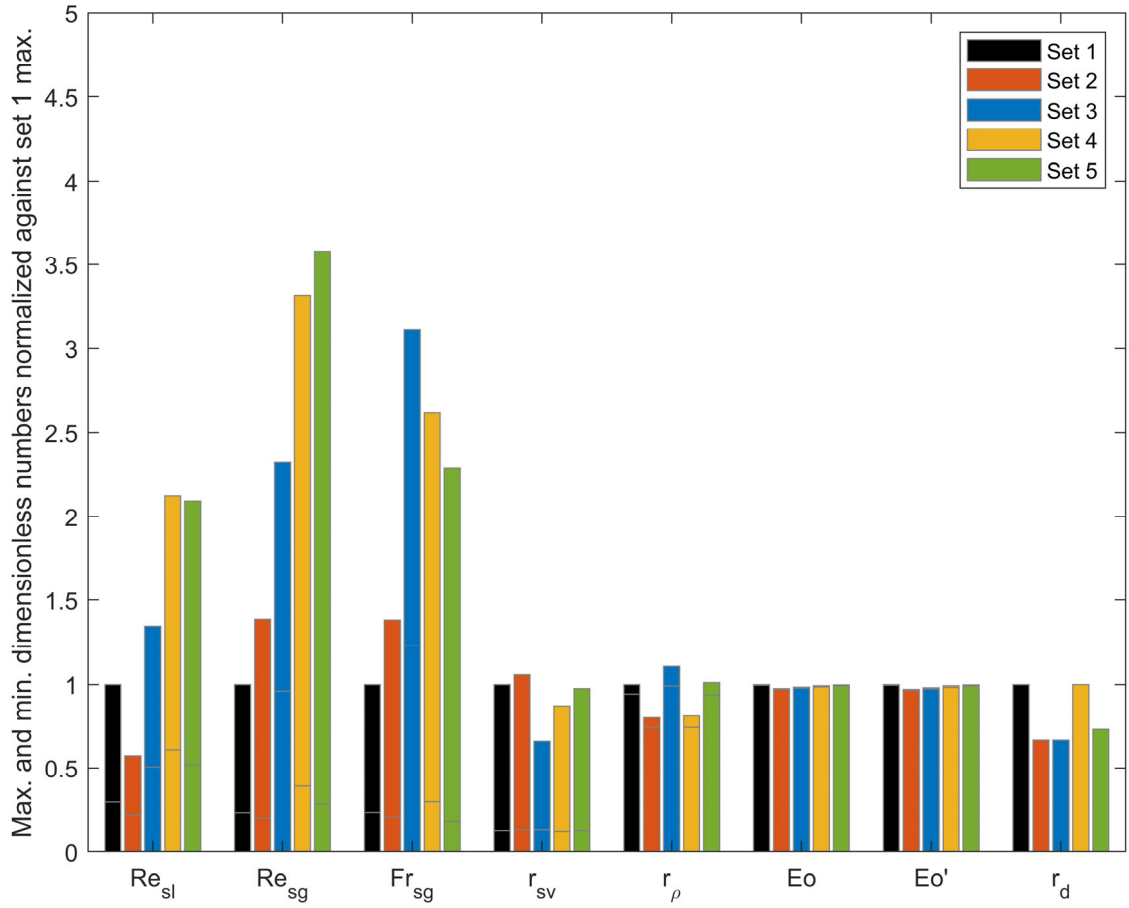

**Figure 9** Normalized range of dimensionless numbers from sets 1–5

**Table 8** Range of dimensionless numbers for all experimental test points

| Dimensionless number | $\text{Re}_{sl}$ | $\text{Re}_{sg}$ | $\text{Fr}_{sg}$ | $r_{l,sv}$ | | $r_{l,\rho}$ | $\text{E}_o$ | $\tilde{\text{E}}_o$ | $r_d$ |
|---|---|---|---|---|---|---|---|---|---|
| Minimum | 29888 | 36595 | 0.50 | 0.33 | (GVF 75%) | 30.74 | 4.61 | 6.09 | 0.07 |
| Maximum | 288199 | 640271 | 8.48 | 2.83 | (GVF 26%) | 45.83 | 4.75 | 6.31 | 0.10 |

### 4.1 Phase fraction

In this Section, the gamma ray equivalent gas fraction, $\alpha_{g,gamma}$, predicted by the $\ln(\text{SL}_{gamma})$ versus $\ln(\text{Fr}_m)$ correlation (with correlation coefficients $a$ and $b$ obtained in Table 7) is compared with the value measured by the gamma ray sensor as follows

$$\alpha_{g,gamma} = \frac{e^{a\ln(\text{Fr}_m)+b}\rho_g\left(\dfrac{\dot{q}_g}{\rho_{VF}A_{MT}}\right)^2 + gD(\rho_{VF}-\rho_l)}{gD(\rho_g-\rho_l)}, \quad (17)$$

where $\dot{q}_g$ is the gas mass flow rate and $A_{MT}$ is the area of MT. Sets 1, 2, and 5 are predicted by their respective coefficients. Sets 3 and 4 are predicted by the correlation coefficients of sets 1 and 2. This is to validate the observation that (i) the pairs that share the same $r_d$, $\text{E}_o$, $\tilde{\text{E}}_o$ have similar gradient, $a$, and (ii) the pairs that share the same $r_{l,\rho}$ have similar offset, $b$. Figure 10 shows the agreement between predicted and measured $\alpha_{g,gamma}$.

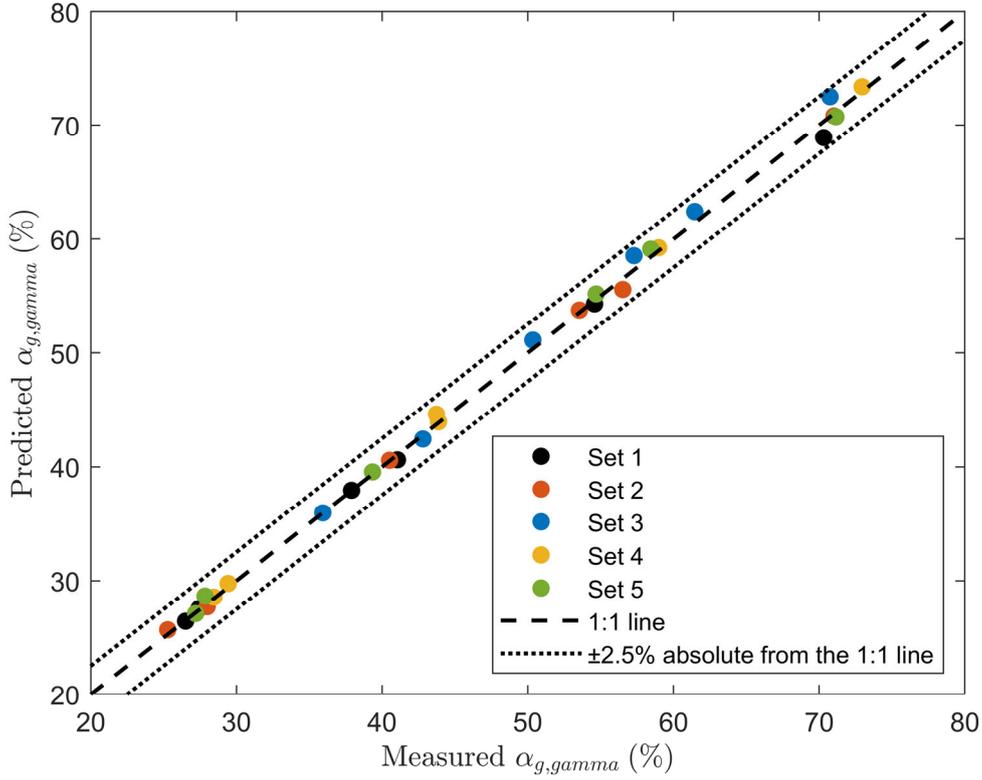

**Figure 10**  Consistency between predicted and measured $\alpha_{g,gamma}$

As can be seen from Figure 10, there is a good agreement between the $\alpha_{g,gamma}$ predicted by $\ln(\text{SL}_{gamma})$ versus $\ln(\text{Fr}_m)$ correlation and the measured $\alpha_{g,gamma}$, with a maximum difference of $\leq 2.5\%$ absolute. Note that the largest difference between the CFD-predicted $\alpha_{g,gamma}$ and that measured by the gamma ray sensor is also $2.5\%$ absolute. Since the $\ln(\text{SL}_{gamma})$ versus $\ln(\text{Fr}_m)$ correlation is obtained from CFD simulations, the difference in $\alpha_{g,gamma}$ predicted by $\ln(\text{SL}_{gamma})$ versus $\ln(\text{Fr}_m)$ correlation cannot be lower than $2.5\%$ absolute. The results also show that good consistency in $\alpha_{g,gamma}$ can be achieved (sets 3 and 4 predictions) using a scaling rule in which the flows with the same $r_d$, $E_o$, and $\tilde{E}_o$ have similar gradient, $a$, and the flows that share the same $r_{l,\rho}$ have similar offset, $b$, in the $\ln(\text{SL}_{gamma})$ versus $\ln(\text{Fr}_m)$ correlation.

### 4.2 Dimensionless Venturi differential pressure

$\Delta P$ can be obtained from

$$\Delta P = \frac{\rho_{g,in} v_{sg,in}^2 (H_{MT} - H_{VI})}{D} \left| \vec{k} \bullet \nabla p \right|. \quad (18)$$

Figure 11 shows the agreement between predicted and measured $\Delta P$.

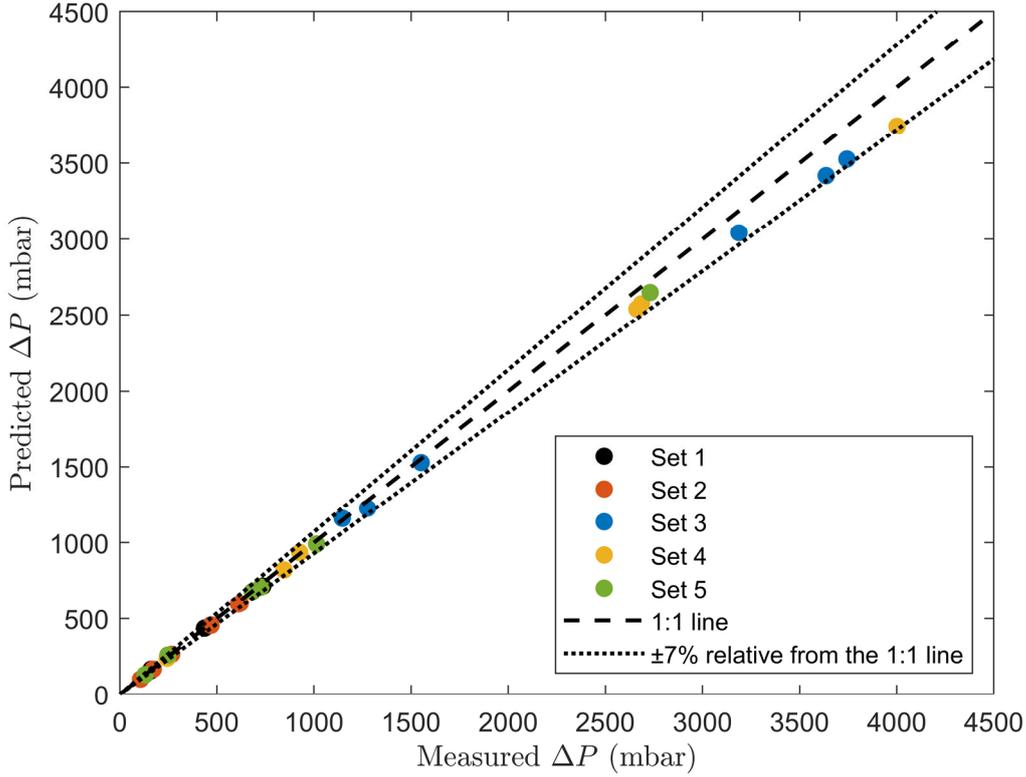

**Figure 11** Consistency between predicted and measured $\Delta P$

From Figure 11, it can be seen that $\Delta P$ predicted by Equation 18 is lower than the measured $\Delta P$, as most of the test points lie below the 1:1 line. This is expected since the energy loss in the Venturi was not fully accounted for; however, four points were observed to have higher predicted $\Delta P$ than the experimental measurements. This may be because in some cases the uncertainty in the $\Delta P$ measurement is greater than the neglected loss. Overall, the agreement between predicted and measured $\Delta P$ is reasonably good, with a difference of $\leq 7\%$ relative to the measured $\Delta P$. Note that $\Delta P$ is used for mesh independence studies. Since Equation 18 shows good prediction of $\Delta P$, the Venturi differential pressure gradient $\left|\vec{k} \bullet \nabla P\right|$ can potentially be used as a criterion indicating mesh independence.

### 4.3 Two-phase discharge coefficient

Since only the gamma ray phase fraction (hence $\rho_m$) and $\Delta P$ are measured in our experiments, Equation 16 cannot be applied to obtain $C_{d,tp}$ experimentally. Hence, $C_{d,tp}$ is defined by

$$C_{d,tp} = \frac{4\dot{q}_m\sqrt{1-\beta^4}}{\pi D^2 \sqrt{2\rho_m\left[\Delta P - \rho_m g\left(H_{MT} - H_{VI}\right)\right]}}, \quad (19)$$

as the two-phase coefficient used to correct the mass flow rate calculated from the measured $\Delta P$ to the reference mass flow rate $\dot{q}_m$. Figure 12 shows the variation of $C_{d,tp}$ with $Re_{sl}$ from experimental measurements.

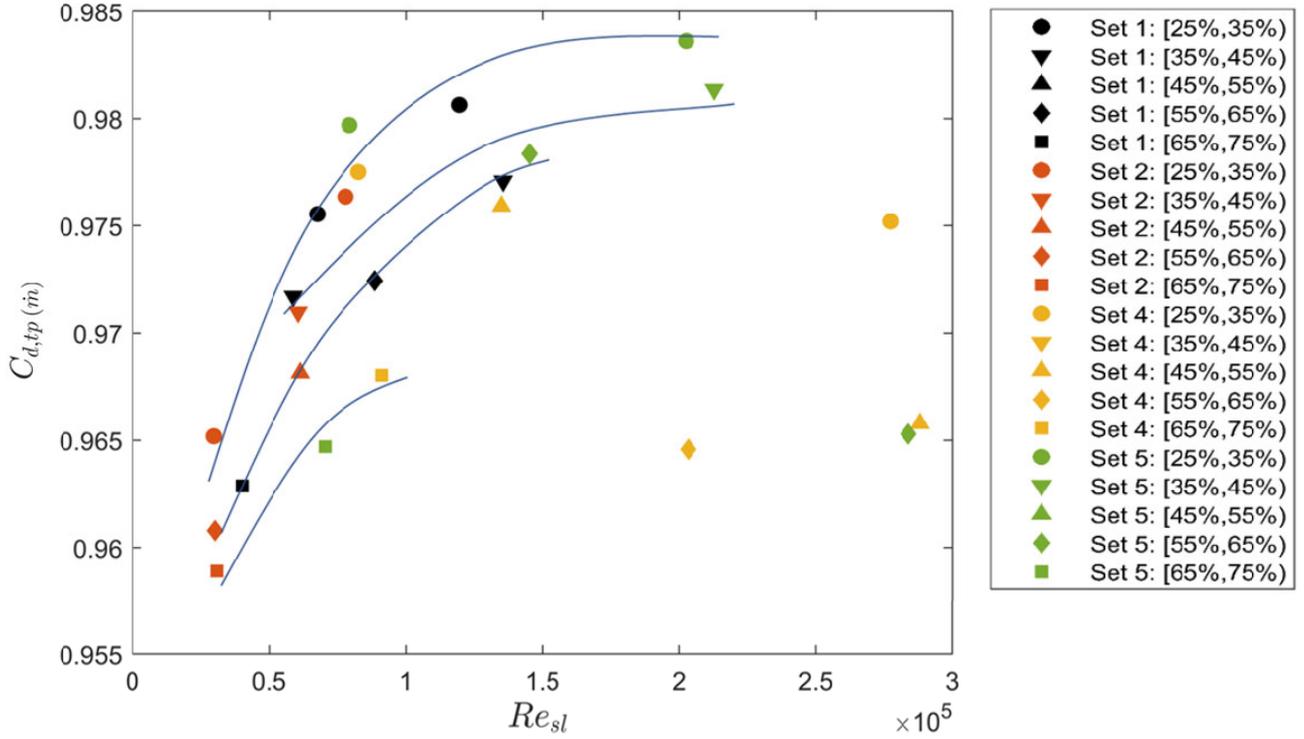

**Figure 12** Variation of $C_{d,tp}$ with $\mathrm{Re}_{sl}$ obtained from flow loop experiments. Lines-of best-fit are shown for the inlet gas volume fraction ranges $[25\%, 35\%)$, $[35\%, 45\%)$, $[45\%, 65\%)$, $[65\%, 70\%)$.

Note that set 3 test points collected in the second flow loop are omitted in Figure 12 because the pipe characteristics related to friction loss may be different from the other sets. As can be seen from Figure 12, the $C_{d,tp}$ versus $\mathrm{Re}_{sl}$ values are clustered according to the inlet GVFs, with lower inlet GVFs having a larger $C_{d,tp}$, which is consistent with the results of the CFD simulation discussed in Section 3.3. However, set 2, which is likely to have different $\mathrm{E}_o$, $\tilde{\mathrm{E}}_o$, and $r_d$ as sets 1, 4, and 5, appears to belong to the same cluster as sets 1, 4, and 5, which are likely to share the same $\mathrm{E}_o$ and $\tilde{\mathrm{E}}_o$. Note that there are a few outliers with $\mathrm{Re}_{sl}$ larger than $2 \times 10^5$ that have lower $C_{d,tp}$ than expected. It is possible that at higher $\mathrm{Re}_{sl}$ (which also indicates higher $\mathrm{Fr}_m$, given the same inlet GVF, fluid property, and pipe size), there may be a change of flow regime where increased gas-liquid interaction may cause greater loss, resulting in lower $C_{d,tp}$.

## 5 Conclusions

In this study, a CFD approach is used to evaluate the performance of scaling rules in gas-liquid two-phase flow in a vertical Venturi in MPFM downstream of a horizontal pipe length. The dimensionless numbers studied, including those associated with flow condition, fluid property, and gas-liquid interaction, are derived from the governing equations, boundary condition for gas-liquid flow, and the dimension of the flow domain. The performance of the scaling rule is evaluated in terms of the phase fraction, the dimensionless Venturi differential pressure, and the two-phase discharge coefficient. The results are validated against the experimental measurements in the flow loop.

Overall, the significance of this work includes: (i) a scaled study of gas-liquid two-phase flow within a vertical Venturi, in which the flow development has never been studied before; (ii) a scaled study of the effect of phase interaction, which is different from previous studies that focused on separated flow. The dimensionless numbers associated with the phase interaction term are found to be related to the phase fraction in the Venturi; (iii) a useful scaling rule has been developed to obtain the coefficient for $\ln(\mathrm{SL}_p)$ versus $\ln(\mathrm{Fr}_m)$ correlation, which is in good agreement with the measured phase fraction. The analytical equation for the dimensionless Venturi differential pressure derived from a set of dimensionless numbers and the $C_{d,tp}$ versus $\mathrm{Re}_{sl}$ relationship can also be used to predict the Venturi differential pressure and friction losses that need to be taken into account for accurate mass flow rate. Once the $\ln(\mathrm{SL}_p)$ versus $\ln(\mathrm{Fr}_m)$ correlation and the $C_{d,tp}$ versus $\mathrm{Re}_{sl}$ relationship are calibrated against the key dimensionless numbers identified, they can be used to potentially reduce the cost and carbon footprint required to perform experiments and CFD simulations, such as the phase fraction, the Venturi differential pressure, and the two-phase discharge coefficient can be predicted by the correlation, given the dimensionless numbers of the gas-liquid flow.

# Appendix

| Set | Test point | Line pressure (bar) | Inlet GVF (%) | Homogeneous inlet velocity (m/s) | Superficial inlet gas velocity (m/s) | Superficial inlet liquid velocity (m/s) | Liquid viscosity (mPa s) | Gas density (kg/m$^3$) | Liquid density (kg/m$^3$) |
|---|---|---|---|---|---|---|---|---|---|
| 1 | 1 | 21.85 | 29.21 | 1.68 | 0.49 | 1.19 | 1.03 | 24.30 | 1005.20 |
|   | 2 | 22.66 | 27.13 | 2.83 | 0.77 | 2.06 | 1.02 | 24.85 | 1006.70 |
|   | 3 | 21.83 | 43.90 | 1.84 | 0.81 | 1.03 | 1.03 | 24.28 | 1005.30 |
|   | 4 | 23.73 | 38.87 | 3.88 | 1.51 | 2.37 | 1.03 | 25.75 | 1005.40 |
|   | 5 | 21.51 | 56.50 | 1.02 | 0.58 | 0.44 | 1.03 | 25.23 | 1005.50 |
|   | 6 | 21.89 | 74.08 | 2.79 | 2.07 | 0.72 | 1.06 | 24.34 | 1002.80 |
| 2 | 1 | 21.64 | 29.42 | 1.46 | 0.43 | 1.03 | 1.62 | 24.10 | 800.31 |
|   | 2 | 23.06 | 26.06 | 3.63 | 0.95 | 2.68 | 1.61 | 25.10 | 800.42 |
|   | 3 | 23.00 | 41.54 | 3.54 | 1.47 | 2.07 | 1.59 | 25.14 | 799.88 |
|   | 4 | 22.04 | 58.70 | 2.50 | 1.47 | 1.03 | 1.60 | 24.39 | 799.97 |
|   | 5 | 23.95 | 54.94 | 4.59 | 2.52 | 2.07 | 1.57 | 25.99 | 798.95 |
|   | 6 | 22.50 | 73.40 | 3.88 | 2.85 | 1.03 | 1.56 | 24.69 | 799.09 |
| 3 | 1 | 19.10 | 58.90 | 10.91 | 6.43 | 4.48 | 1.60 | 18.34 | 801.82 |
|   | 2 | 19.20 | 51.40 | 10.68 | 5.49 | 5.19 | 1.58 | 17.90 | 801.08 |
|   | 3 | 18.96 | 36.10 | 9.52 | 3.44 | 6.08 | 1.56 | 17.46 | 800.37 |
|   | 4 | 18.11 | 42.90 | 5.91 | 2.54 | 3.37 | 1.53 | 19.23 | 800.26 |
|   | 5 | 18.44 | 63.10 | 8.13 | 5.13 | 3.00 | 1.52 | 19.27 | 799.79 |
|   | 6 | 18.21 | 73.30 | 8.24 | 6.04 | 2.20 | 1.49 | 19.44 | 798.15 |
| 4 | 1 | 29.18 | 29.92 | 2.06 | 0.62 | 1.44 | 1.03 | 32.20 | 1005.10 |
|   | 2 | 30.07 | 30.00 | 6.88 | 2.06 | 4.82 | 1.02 | 30.50 | 1005.80 |
|   | 3 | 29.69 | 45.00 | 4.38 | 1.97 | 2.41 | 1.05 | 32.10 | 1003.40 |
|   | 4 | 30.81 | 44.97 | 9.39 | 4.22 | 5.17 | 1.05 | 29.83 | 1002.70 |
|   | 5 | 30.74 | 59.89 | 9.01 | 5.40 | 3.61 | 1.04 | 31.19 | 1003.50 |
|   | 6 | 30.17 | 74.96 | 6.72 | 5.04 | 1.68 | 1.08 | 32.48 | 999.88 |
| 5 | 1 | 23.27 | 27.60 | 2.23 | 0.62 | 1.61 | 1.02 | 25.00 | 1005.30 |
|   | 2 | 24.34 | 40.38 | 3.14 | 1.27 | 1.87 | 1.04 | 25.84 | 1003.20 |
|   | 3 | 23.70 | 56.87 | 4.10 | 2.33 | 1.77 | 1.06 | 25.49 | 1000.70 |
|   | 4 | 22.33 | 74.27 | 3.91 | 2.90 | 1.01 | 1.03 | 24.44 | 1004.60 |
|   | 5 | 21.81 | 30.02 | 1.66 | 0.50 | 1.16 | 1.04 | 23.91 | 1002.40 |
|   | 6 | 24.56 | 59.98 | 4.14 | 2.48 | 1.66 | 1.04 | 24.07 | 1002.70 |

# Acknowledgements

This research is supported by Singapore Economic Development Board, Singapore Ministry of Education Academic Research Fund Tier 1 (RG75/20) and SLB (formerly Schlumberger).